\documentstyle[11pt,newpasp,epsf,twoside]{article}
\def\edcomment#1{\iffalse\marginpar{\raggedright\sl#1\/}\else\relax\fi}
\reversemarginpar

\begin{document}
\title{A Model of the Temporal Variability of Optical Light from Extrasolar Terrestrial Planets}
\author{Eric B.\  Ford}
\affil{Department of Astrophysical Sciences, Princeton University, Peyton Hall - Ivy Lane, Princeton, NJ 08544}
\author{Sara Seager}
\affil{Institute for Advanced Study, Einstein Drive, Princeton, NJ 08540}
\author{Edwin L.\ Turner}
\affil{Department of Astrophysical Sciences, Princeton University, Peyton Hall - Ivy Lane, Princeton, NJ 08544}

\begin{abstract}
The light scattered by an extrasolar Earth-like planet's surface and
atmosphere will vary in intensity and color as the planet rotates; the
resulting light curve will contain information about the planet's
properties.  Since most of the light comes from a small fraction of
the planet's surface, the temporal flux variability can be quite
significant, $\sim 10-100\%$.  In addition, for cloudless Earth-like
extrasolar planet models, qualitative changes to the surface (such as
ocean fraction, ice cover) significantly affect the light curve.
Clouds dominate the temporal variability of the Earth but can be
coherent over several days.  In contrast to Earth's temporal
variability, a uniformly, heavily clouded planet (e.g.\ Venus), would
show almost no flux variability. We present light curves for an
unresolved Earth and for Earth-like model planets calculated by
changing the surface features. This work suggests that meteorological
variability and the rotation period of an Earth-like planet could be
derived from photometric observations. The inverse problem of deriving
surface properties from a given light curve is complex and will
require much more investigation.
\end{abstract}

Terrestrial planets around nearby stars are of enormous
interest, especially any that orbit in habitable zones (surface
conditions compatible with liquid water), since they might have global
environments similar to Earth's and even harbor life.
NASA and ESA are now planning very challenging and ambitious space
missions-Terrestrial Planet Finder and Darwin respectively-to detect
and characterize terrestrial planets orbiting nearby Sun-like stars.
Very different designs are being considered at both optical and mid-IR
wavelengths, but all have the goal of spectroscopic characterization
of the atmospheric composition including the capability to detect gases
important for or caused by life on Earth (e.g.\ O$_2$, O$_3$, CO$_2$,
CH$_4$ and H$_2$O). A mission capable of measuring these spectral
features would have the signal-to-noise necessary to
measure photometric variability of the unresolved planet. Photometry
used to investigate a planet in less integration time than necessary
for spectroscopy or could be done concurrently.

\begin{figure}[phtb]
\plotone{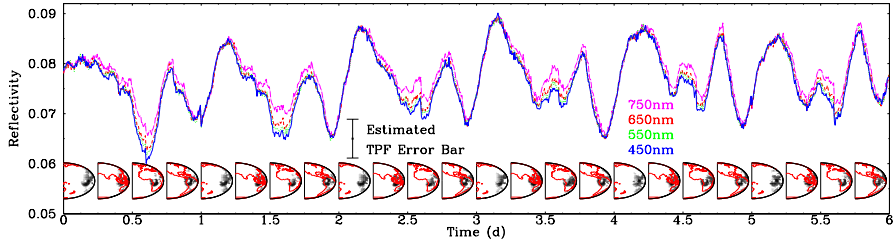}
\caption{Model light curve for Earth.
The color of each line corresponds to the wavelength (450, 550, 650,
750~nm).  The pictures below show the viewing geometry (continents
outlined in red) and the amount of light contributed from each part of
the surface (dark regions make largest contribution).  Since cloud
patterns produce the largest features in the light curve and are
coherent over several days, the rotation period can determined and a
daily average light curve can be calculated.  Monte Carlo statistics
produce the small ``noise''.  The small vertical jumps at three hour
intervals is due to using discrete cloud cover maps.  }
\end{figure}

In Fig.\ 1, we show the light curve of our Earth model for six
consecutive days using cloud coverage obtained from satellite data.
Photometric variations on the order of 20\% could be easily detected
by a TPF capable of spectroscopy.  

Observations of the dark side of the Moon can be used to measure the
reflectivity of the Earth (Goode et al.\ 2001).  Fig.\ 2 (right) shows
the viewing geometetry and illustrates why Earthshine observations are
limited to certain viewing angles and times of day.  Our model
accurately reproduces both the mean reflectivity and the degree of
variability (error bar at $90^\circ$ represents 1$\sigma$ variance
between realizations) for widely separated days (See Fig.\ 2 left).
The difference in magnitude of an extrasolar planet is given by
\begin{equation}
\Delta m = 22.60 + 5 \log \left(\frac{r}{AU}\right) - 5 \log \left(\frac{R_p}{R_\oplus}\right) - 2.5 \log
\left(\mathcal{R}\right),
\end{equation}
where $r$ is the star-planet distance, $R_p$ is the planet radius, and
$\mathcal{R}$ is the reflectivity which varies with phase angle and
viewing geometry.

\begin{figure}[pthb]
\plottwo{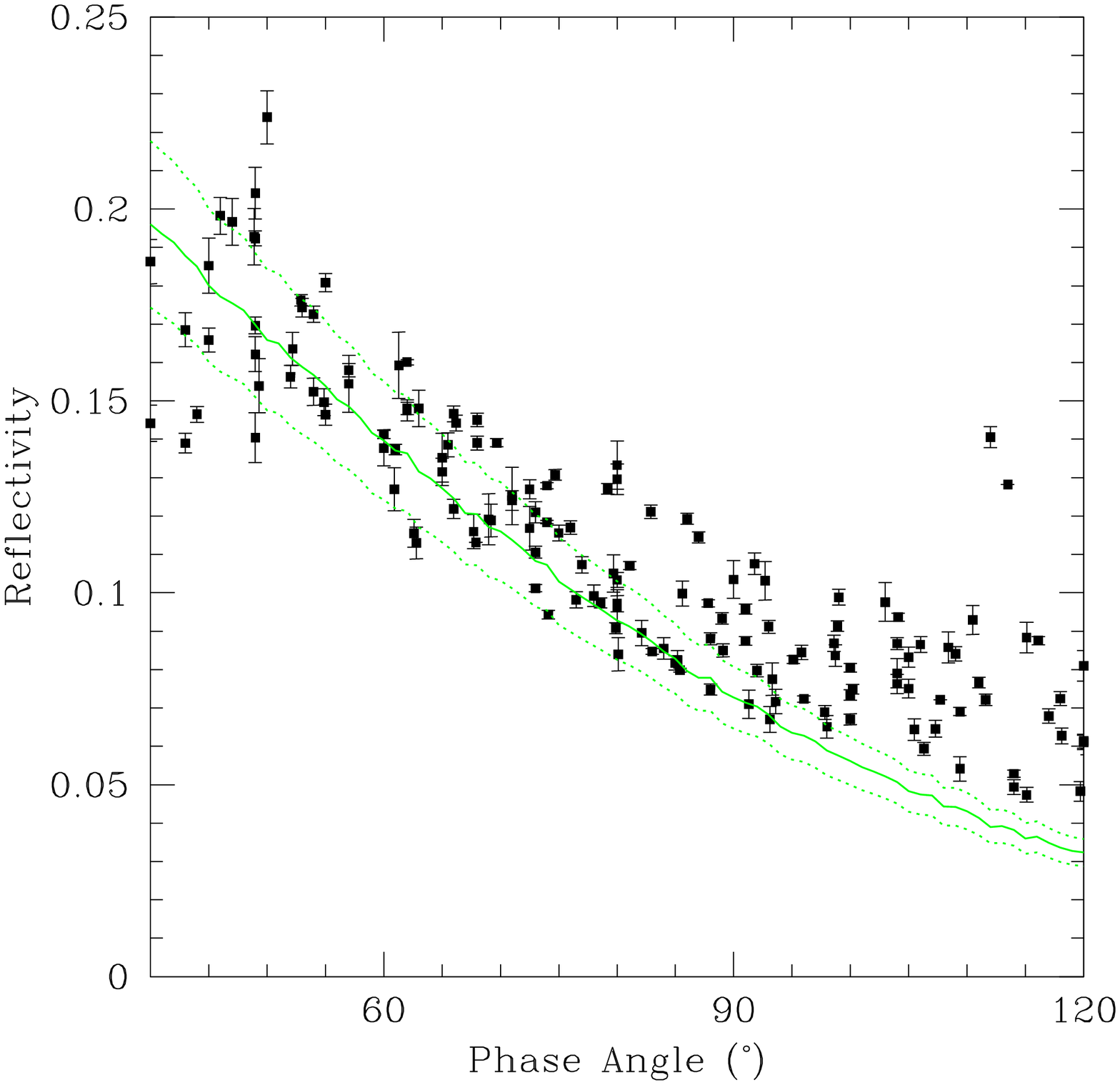}{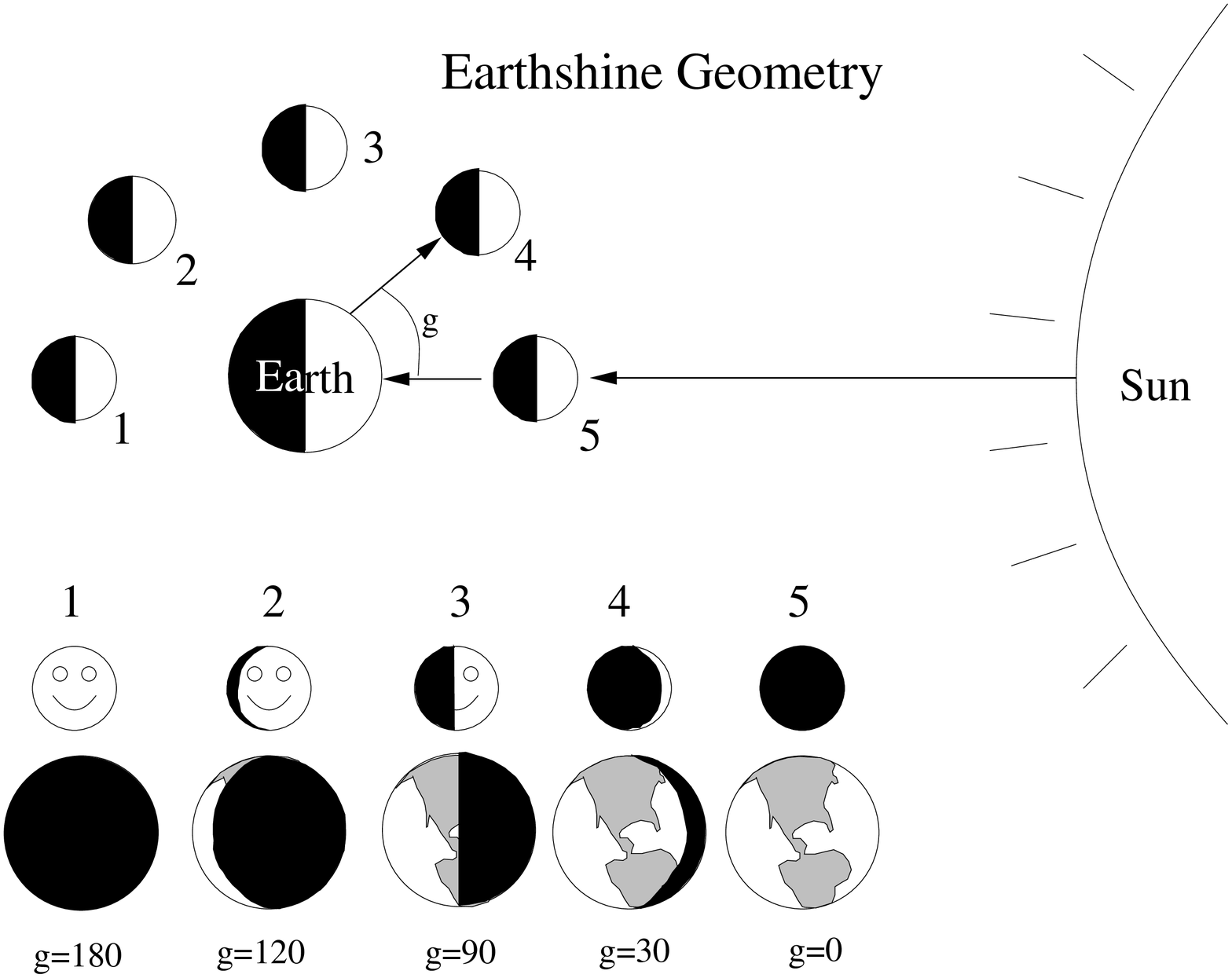}
\caption{Left: Comparison of model to Earthshine observations.  Here
we compare our Earth model (line) to the observations (points) taken
near $90^\circ$ phase angle.
Right: Viewing geometry for Earthshine observations.  
}
\end{figure}

The spectrum shown in Fig.\ 3 (left) illustrates a dramatic
rise in reflectivity from the optical to the near-IR1 (see Fig.\
3 left).  Remote sensing satellites routinely use this feature to
recognize vegetation on the Earth.  The high reflectance at near-IR
wavelengths is due to the arrangement of cells and air gaps in the
leaves and is believed to allow plants to absorb light useful for
photosynthesis while reflecting light which would only produce
destructive excess heat.

We have used our Earth model to calculate the variation of Earth's
color using the actual distribution of clouds from satellite data and
theoretical spectra from Traub \& Jucks (2002) (see Fig.\ 3 right).
The greater variability of the color centered on the red edge may be
recognizable in Earthshine observations.

\begin{figure}[pbht]
\plottwo{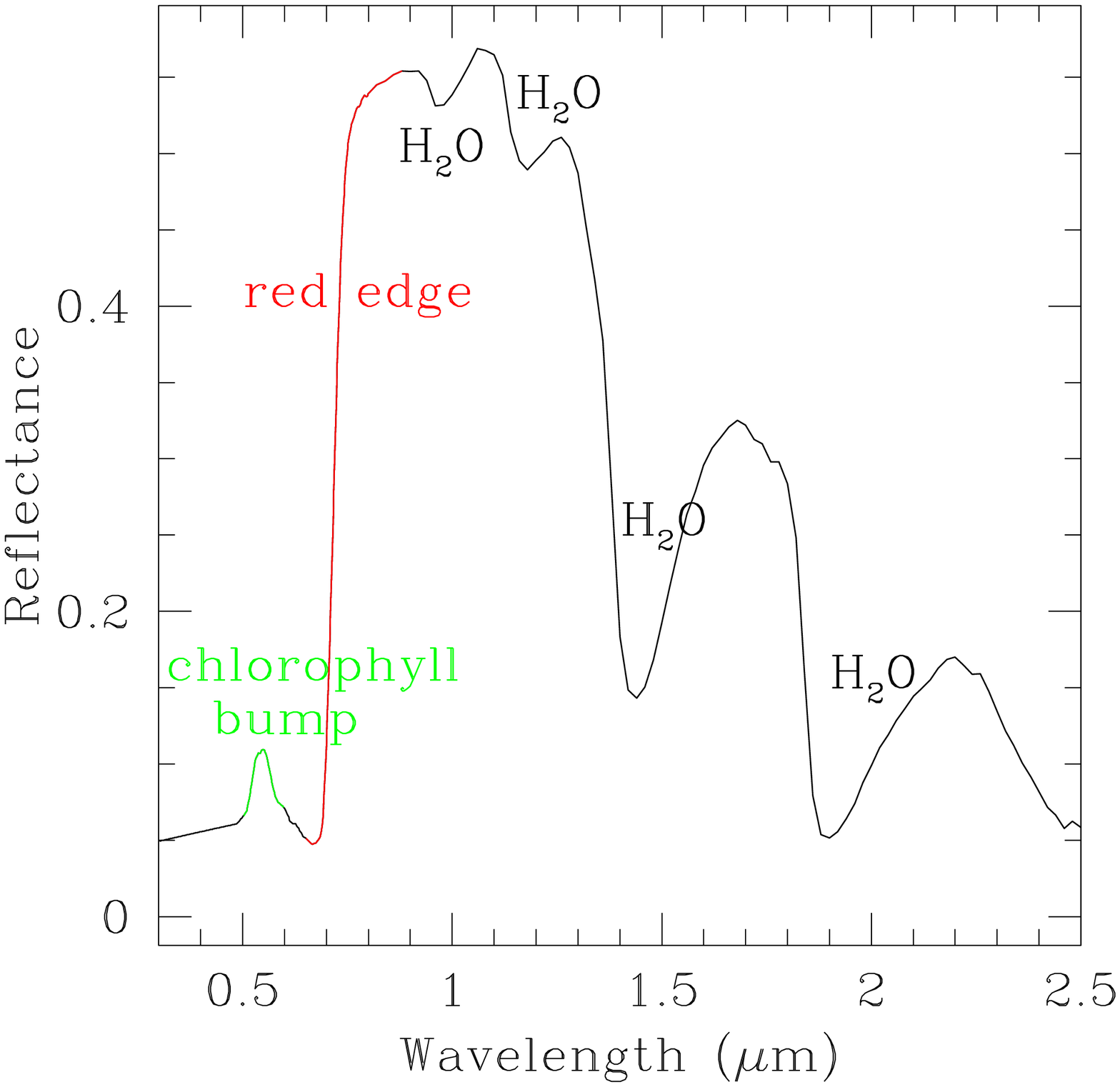}{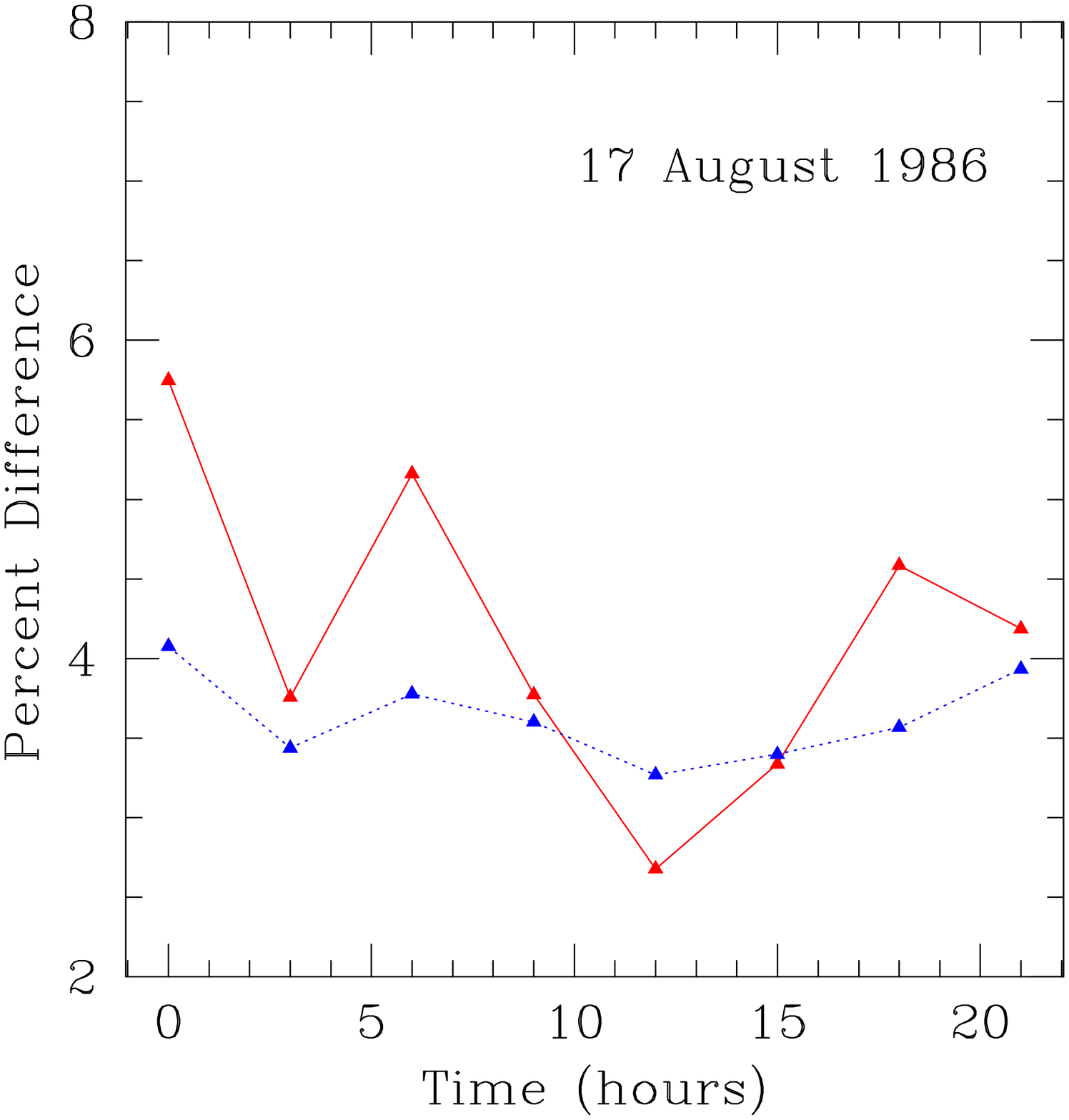}
\caption{Left: Spectrum of a deciduous leaf.
The small bump comes from chlorophyll absorption bands on either side
of 550nm and causes people to perceive plants as green.  The much
larger sharp rise (near 800nm, slightly beyond the optical) is known
as the ``red-edge''.
Right: Variability of Earth's color.
The solid red line shows a color
$\left(\frac{I(750-800\mathrm{nm}) -
I(700-650\mathrm{nm})}{I(750-800\mathrm{nm})}\right)$ chosen to
emphasize variability due to vegetation's red edge.  For comparison,
the dotted blue line shows a color $\left(\frac{I(850-800\mathrm{nm})
- I(750-800\mathrm{nm})}{I(750-800\mathrm{nm})}\right)$ which is less
sensitive to vegetation.  
}
\end{figure}

As the Earth rotates, different features rotate into view, causing
significant variations in the total light from the entire planet.
Once a rotation period is measured, observations over many rotation
periods could be folded to obtain average light curves for summer,
winter, and the entire year (see Fig.\ 4 left).  We considered
plausible cloudless Earth-like planets by altering the surface map of
the Earth (see Fig.\ 4 right).  While many of these qualitative
surface changes result in significant changes to the light curves,
deducing the surface of an extrasolar planet from its light curve
could be quite difficult.

\begin{figure}[htb]
\plottwo{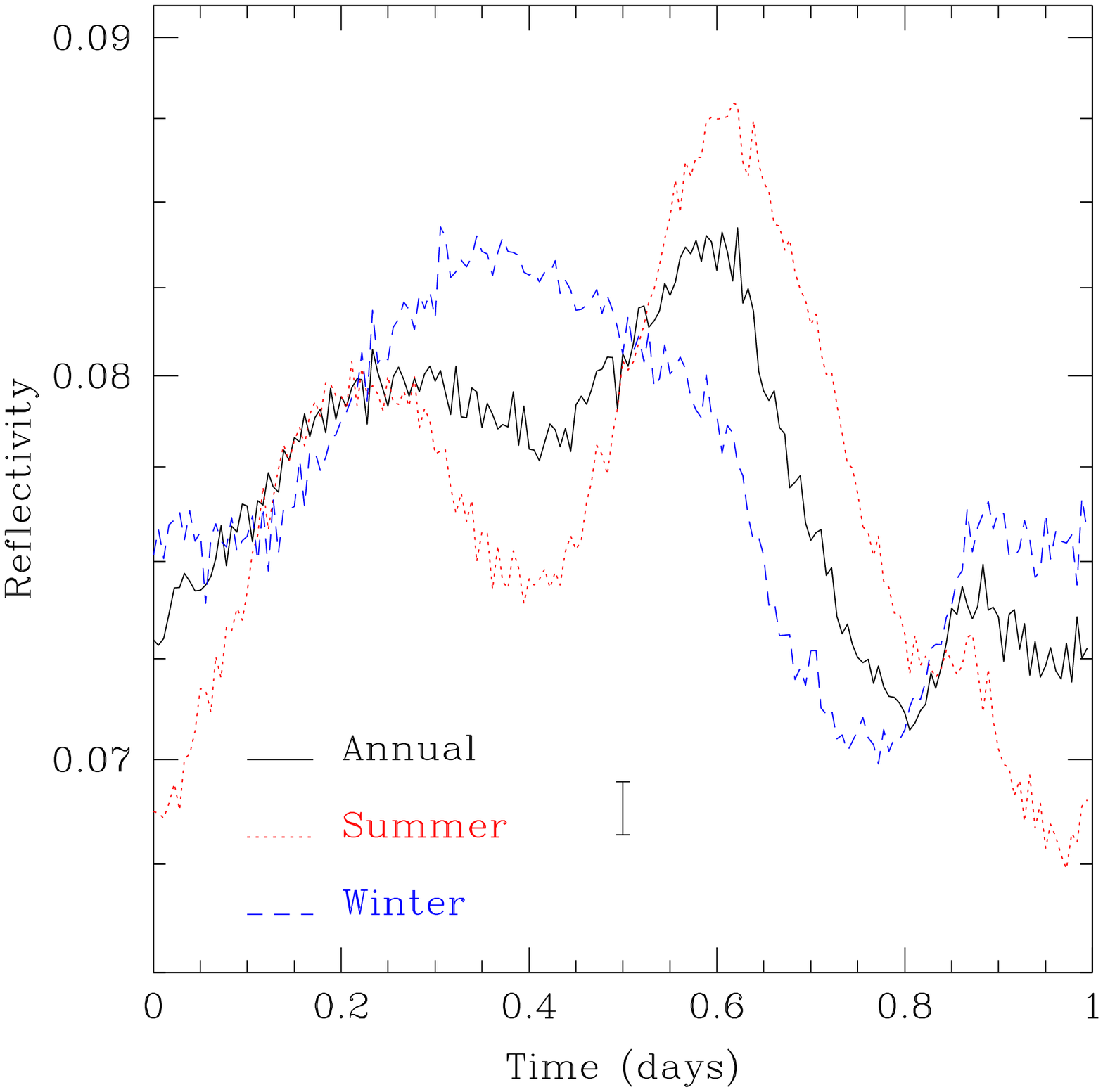}{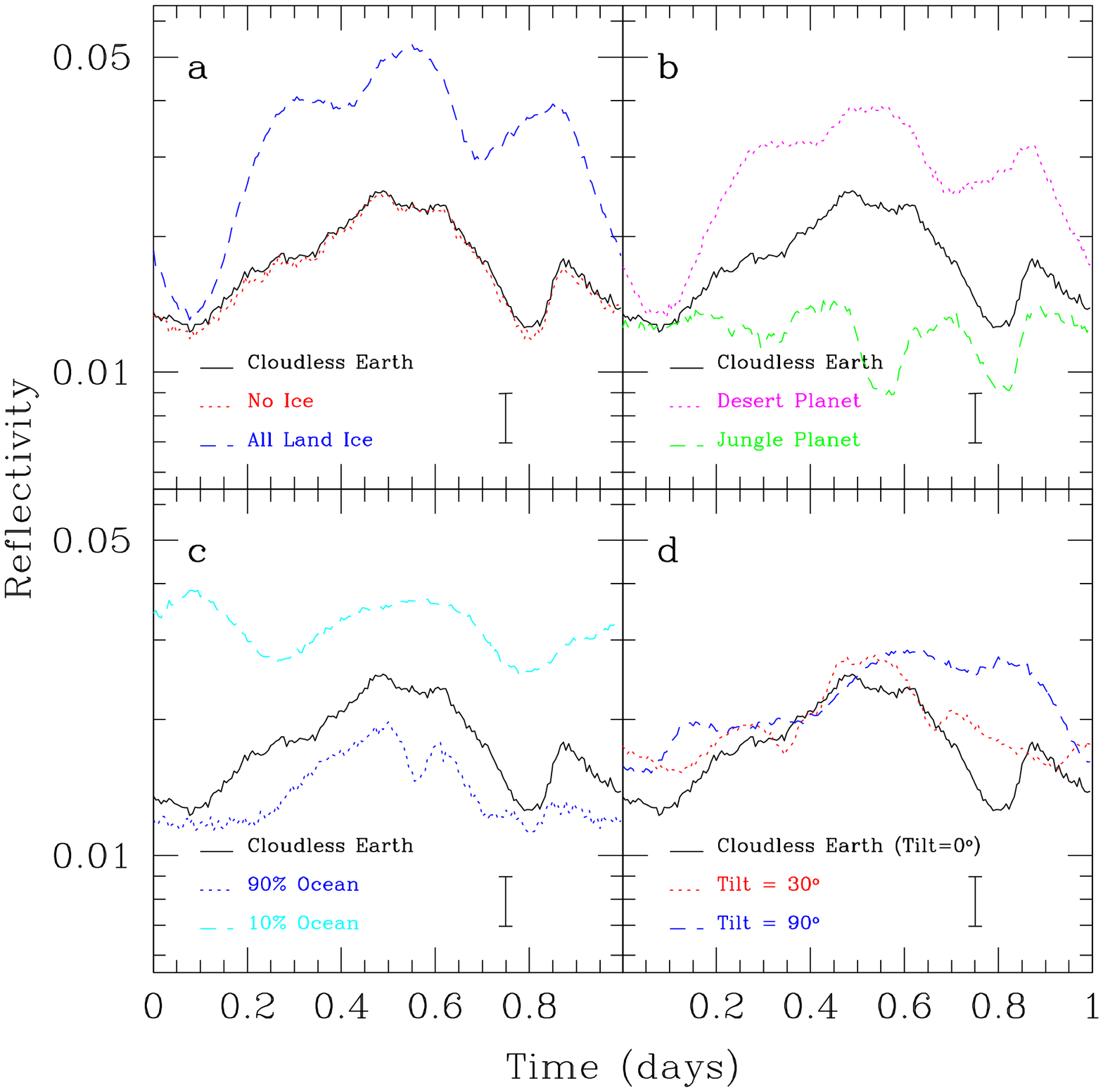}
\caption{Left: Model seasonal average light curves of cloudy Earth.
Here the
differences between the light curves are solely due to seasonal changes
in the average cloud conditions, since changes in the snow cover is
not included in the model.  Detecting seasonal changes in a planet's
light curve could provide information about seasons.  Interpretation
could be complicated by changes in the viewing geometry.
Right: Model for cloudless Earth-like planets.
We consider plausible Earth-like planets by altering the
surface map of the Earth.  In panel {\bf a} all the land is covered
with ice or all the ice is replaced with desert.  In
panel {\bf b} all the land is covered with thick forest or
desert.  In panel {\bf c} the fraction of the surface
covered with oceans is varied.  All the above light curves have
significant variability, $\sim10-100\%$.  Since qualitative
changes to the surface significantly affect the light curve, the light
curve may constrain surface properties.  In panel {\bf d} we
consider the cloudless Earth for different obliquities.  
}
\end{figure}

\end{document}